\documentclass[preprint,superscriptaddress,nofootinbib]{revtex4}
  \usepackage{graphicx,subfigure}
  
\begin{document}
  \title{The ${\Upsilon}(nS)$ ${\to}$ $B_{c}^{\ast}{\pi}$,
         $B_{c}^{\ast}K$ decays with perturbative QCD approach}
  \author{Junfeng Sun}
  \affiliation{Institute of Particle and Nuclear Physics,
              Henan Normal University, Xinxiang 453007, China}
  \author{Yueling Yang}
  \affiliation{Institute of Particle and Nuclear Physics,
              Henan Normal University, Xinxiang 453007, China}
  \author{Qin Chang}
  \affiliation{Institute of Particle and Nuclear Physics,
              Henan Normal University, Xinxiang 453007, China}
  \author{Gongru Lu}
  \affiliation{Institute of Particle and Nuclear Physics,
              Henan Normal University, Xinxiang 453007, China}
  \author{Jinshu Huang}
  \affiliation{College of Physics and Electronic Engineering,
              Nanyang Normal University, Nanyang 473061, China}

  \begin{abstract}
  Besides the traditional strong and electromagnetic decay modes,
  ${\Upsilon}(nS)$ meson can also decay through the weak
  interactions within the standard model of elementary particle.
  With anticipation of copious ${\Upsilon}(nS)$ data samples at the
  running LHC and coming SuperKEKB experiments, the two-body
  nonleptonic bottom-changing ${\Upsilon}(nS)$ ${\to}$
  $B_{c}^{\ast}{\pi}$, $B_{c}^{\ast}K$ decays ($n$ $=$ 1, 2, 3)
  are investigated with perturbative QCD approach firstly.
  The absolute branching ratios for ${\Upsilon}(nS)$ ${\to}$ $B_{c}^{\ast}{\pi}$
  and $B_{c}^{\ast}K$ decays are estimated to reach up to about $10^{-10}$
  and $10^{-11}$, respectively, which might possibly be measured
  by the future experiments.
  \end{abstract}
  \maketitle

  \section{Introduction}
  \label{sec01}
  The upsilon ${\Upsilon}(nS)$ meson is the spin-triplet $S$-wave
  state of bottomonium (bound state consisting of bottom
  quark $b$ and anti-bottom quark $\bar{b}$) with well-established
  quantum number of $I^{G}J^{PC}$ $=$ $0^{-}1^{--}$ \cite{pdg}.
  The characteristic narrow decay widths of ${\Upsilon}(nS)$ mesons
  for $n$ $=$ 1, 2 and 3 provide insight into the study of strong
  interactions. [see Table. \ref{tab:bb}, and note that for
  simplicity, ${\Upsilon}(nS)$ will denote ${\Upsilon}(1S)$,
  ${\Upsilon}(2S)$ and ${\Upsilon}(3S)$ mesons in the following
  content if not specified definitely.]
  The mass of ${\Upsilon}(nS)$ meson is below the $B$ meson pair
  threshold.
  The ${\Upsilon}(nS)$ meson decays into bottomed hadrons
  through strong and electromagnetic interactions are
  forbidden by the law of conservation of flavor number.
  The bottom-changing ${\Upsilon}(nS)$ decays can occur
  only via the weak interactions within the standard model,
  although with tiny incidence probability.
  Both constituent quarks of upsilons can decay
  individually, which provide an alternative system for
  investigating the weak decay of heavy-flavored hadrons.
  In this paper, we will study the nonleptonic ${\Upsilon}(nS)$
  ${\to}$ $B_{c}^{\ast}P$ ($P$ $=$ ${\pi}$ and $K$) weak decays with
  perturbative QCD (pQCD) approach \cite{pqcd1,pqcd2,pqcd3}.

   \begin{table}[h]
   \caption{Summary of mass, decay width, on(off)-peak luminosity and numbers
   of ${\Upsilon}(nS)$.}
   \label{tab:bb}
   \begin{ruledtabular}
   \begin{tabular}{ccccccc}
    & \multicolumn{2}{c}{properties \cite{pdg}}
    & \multicolumn{2}{c}{luminosity ($fb^{-1}$) \cite{epjc74}}
    & \multicolumn{2}{c}{numbers ($10^{6}$) \cite{epjc74}}
      \\ \cline{2-3} \cline{4-5} \cline{6-7}
      meson & mass (MeV) & width (keV) & Belle & BaBar & Belle & BaBar \\ \hline
   ${\Upsilon}(1S)$ & $9460.30{\pm}0.26$ & $54.02{\pm}1.25$
                    & 5.7 (1.8) & ......
                    & $102{\pm}2$ & ......   \\
   ${\Upsilon}(2S)$ & $10023.26{\pm}0.31$ & $31.98{\pm}2.63$
                    & 24.9 (1.7) & 13.6 (1.4)
                    & $158{\pm}4$ & $98.3{\pm}0.9$ \\
   ${\Upsilon}(3S)$ & $10355.2{\pm}0.5 $ & $20.32{\pm}1.85$
                    & 2.9 (0.2) & 28.0 (2.6)
                    & $11{\pm}0.3$ &  $121.3{\pm}1.2$
   \end{tabular}
   \end{ruledtabular}
   \end{table}

  Experimentally, (1) over $10^{8}$ ${\Upsilon}(nS)$ data samples
  have been accumulated at Belle and BaBar experiments \cite{epjc74}.
  More and more upsilon data samples will be collected at the running
  hadron collider LHC and the forthcoming $e^{+}e^{-}$ collider
  SuperKEKB\footnotemark[1].
  \footnotetext[1]{The SuperKEKB has started commissioning test run
  (http://www.kek.jp/en/NewsRoom/Release).}
  There seems to exist a realistic possibility to explore
  ${\Upsilon}(nS)$ weak decay at future experiments.
  (2) Signals of the ${\Upsilon}(nS)$ ${\to}$ $B_{c}^{\ast}{\pi}$,
  $B_{c}^{\ast}K$ decays should be easily distinguished
  with ``charge tag'' technique,
  due to the facts that the back-to-back final states with
  different electric charges have definite momentum and energy
  in the rest frame of ${\Upsilon}(nS)$ meson.
  (3) The $B_{c}^{\ast}$ meson has not been observed experimentally
  by now. The $B_{c}^{\ast}$ meson production via the
  strong interaction are suppressed due to the simultaneous presence
  of two heavy quarks with different flavors and higher order in QCD
  coupling constant ${\alpha}_{s}$.
  The ${\Upsilon}(nS)$ ${\to}$ $B_{c}^{\ast}{\pi}$, $B_{c}^{\ast}K$
  decays provide a novel pattern to study the $B_{c}^{\ast}$ meson
  production.
  The identification of a single explicitly flavored
  $B_{c}^{\ast}$ meson could be used as an effective selection
  criterion to detect upsilon weak decays.
  Moreover, the radiative decay of $B_{c}^{\ast}$ meson
  provide a useful extra signal and a powerful constraint\footnotemark[2].
  Of course, any discernible evidences of an anomalous production rate
  of single bottomed meson from upsilon decays might be a
  hint of new physics.
  \footnotetext[2]{The investigation on the radiative decay of
  $B_{c}^{\ast}$ meson can be found in,  for example, Ref. \cite{epjc73}
  with QCD sum rules.}

  Theoretically, many attractive QCD-inspired methods have been
  developed recently to describe the exclusive nonleptonic decay
  of heavy-flavored mesons,
  such as the pQCD approach \cite{pqcd1,pqcd2,pqcd3},
  the QCD factorization approach \cite{qcdf1,qcdf2,qcdf3},
  soft and collinear effective theory \cite{scet1,scet2,scet3,scet4},
  and have been applied widely to vindicate
  measurements on $B$ meson decays.
  The upsilon weak decay permits one to further constrain parameters
  obtained from $B$ meson decay, and cross comparisons provide an
  opportunity to test various phenomenological models.
  The upsilon weak decay possess a unique structure due to the
  Cabibbo-Kobayashi-Maskawa (CKM) matrix properties which predicts
  the channels with one $B_{c}^{(\ast)}$ meson are dominant.
  The ${\Upsilon}(nS)$ ${\to}$ $B_{c}^{\ast}P$ decay belongs to
  the favorable $b$ ${\to}$ $c$ transition, which should, in
  principle, have relatively large branching ratio among upsilon
  weak decays. However,
  there is still no theoretical study devoted to the
  ${\Upsilon}(nS)$ ${\to}$ $B_{c}^{\ast}P$ decay for the moment.
  In this paper, we will present a phenomenological
  investigation on ${\Upsilon}(nS)$ ${\to}$ $B_{c}^{\ast}P$
  weak decay with the pQCD approach to supply a ready
  reference for the future experiments.

  This paper is organized as follows.
  Section \ref{sec02} focus on theoretical framework
  and decay amplitudes for ${\Upsilon}(nS)$ ${\to}$
  $B_{c}^{\ast}{\pi}$, $B_{c}^{\ast}K$ weak decays.
  Section \ref{sec03} is devoted to numerical results
  and discussion. The last section is a summary.

  \section{theoretical framework}
  \label{sec02}
  \subsection{The effective Hamiltonian}
  \label{sec0201}
  Theoretically, the ${\Upsilon}(nS)$ ${\to}$ $B_{c}^{\ast}{\pi}$,
  $B_{c}^{\ast}K$ weak decays are described by an effective
  bottom-changing Hamiltonian based on operator product
  expansion \cite{9512380}:
   \begin{equation}
  {\cal H}_{\rm eff}\ =\
   \frac{G_{F}}{\sqrt{2}}\,
   \sum\limits_{q=d,s} V_{cb} V_{uq}^{\ast}
   \Big\{ C_{1}({\mu})\,O_{1}({\mu})
  +C_{2}({\mu})\,O_{2}({\mu}) \Big\}
   + {\rm h.c.}
   \label{hamilton},
   \end{equation}
  where $G_{F}$ ${\simeq}$ $1.166{\times}10^{-5}\,{\rm GeV}^{-2}$
  \cite{pdg} is the Fermi coupling constant;
  the CKM factors $V_{cb}V_{ud}^{\ast}$ and $V_{cb}V_{us}^{\ast}$
  correspond to ${\Upsilon}(nS)$ ${\to}$ $B_{c}^{\ast}{\pi}$
  and $B_{c}^{{\ast}}K$ decays, respectively;
  with the Wolfenstein parameterization, the CKM factors
  are expanded as a power series in a small Wolfenstein
  parameter ${\lambda}$ ${\sim}$ $0.2$ \cite{pdg}:
  \begin{eqnarray}
  V_{cb}V_{ud}^{\ast} &=&
               A{\lambda}^{2}
  - \frac{1}{2}A{\lambda}^{4}
  - \frac{1}{8}A{\lambda}^{6}
  +{\cal O}({\lambda}^{7})
  \label{eq:ckm-pi},  \\
  V_{cb}V_{us}^{\ast} &=& A{\lambda}^{3}
  +{\cal O}({\lambda}^{7})
  \label{eq:ckm-k}.
  \end{eqnarray}
  The local tree operators $Q_{1,2}$ are defined as:
    \begin{eqnarray}
    O_{1} &=&
  [ \bar{c}_{\alpha}{\gamma}_{\mu}(1-{\gamma}_{5})b_{\alpha} ]
  [ \bar{q}_{\beta} {\gamma}^{\mu}(1-{\gamma}_{5})u_{\beta} ]
    \label{q1}, \\
    O_{2} &=&
  [ \bar{c}_{\alpha}{\gamma}_{\mu}(1-{\gamma}_{5})b_{\beta} ]
  [ \bar{q}_{\beta}{\gamma}^{\mu}(1-{\gamma}_{5})u_{\alpha} ]
    \label{q2},
    \end{eqnarray}
  where ${\alpha}$ and ${\beta}$ are color indices and
  the sum over repeated indices is understood.

  The scale ${\mu}$ factorizes physics contributions into
  short- and long-distance dynamics.
  The Wilson coefficients $C_{i}(\mu)$ summarize the physics
  contributions at scale higher than ${\mu}$, and are
  calculable with the renormalization group improved
  perturbation theory.
  The hadronic matrix elements (HME), where
  the local operators are inserted between initial and final
  hadron states, embrace the physics contributions below
  scale of ${\mu}$.
  To obtain decay amplitudes, the remaining work
  is to calculate HME properly by separating from
  perturbative and nonperturbative contributions.

  \subsection{Hadronic matrix elements}
  \label{sec0202}
  Based on Lepage-Brodsky approach for exclusive processes \cite{prd22},
  HME is commonly expressed as a convolution integral of hard scattering
  subamplitudes containing perturbative contributions with universal
  wave functions reflecting nonperturbative contributions.
  In order to effectively regulate endpoint singularities and provide a
  naturally dynamical cutoff on nonperturbative contributions, transverse
  momentum of valence quarks is retained and the Sudakov factor is
  introduced within the pQCD framework \cite{pqcd1,pqcd2,pqcd3}.
  Phenomenologically, the pQCD's decay amplitude could be divided into
  three parts: the Wilson coefficients $C_{i}$ incorporating the
  hard contributions above typical scale of $t$, process-dependent
  rescattering subamplitudes $T$ accounting for the heavy quark decay,
  and wave functions ${\Phi}$ of all participating hadrons,
  which is expressed as
  \begin{equation}
  {\int} dk\,
  C_{i}(t)\,T(t,k)\,{\Phi}(k)e^{-S}
  \label{hadronic},
  \end{equation}
  where $k$ is the momentum of valence quarks, and
  $e^{-S}$ is the Sudakov factor.

  \subsection{Kinematic variables}
  \label{sec0203}
  The light cone kinematic variables in the ${\Upsilon}(nS)$
  rest frame are defined as follows.
  \begin{equation}
  p_{\Upsilon}\, =\, p_{1}\, =\, \frac{m_{1}}{\sqrt{2}}(1,1,0)
  \label{kine-p1},
  \end{equation}
  \begin{equation}
  p_{B_{c}^{\ast}}\, =\, p_{2}\, =\, (p_{2}^{+},p_{2}^{-},0)
  \label{kine-p2},
  \end{equation}
  \begin{equation}
  p_{3}\, =\, (p_{3}^{-},p_{3}^{+},0)
  \label{kine-p3},
  \end{equation}
  \begin{equation}
  p_{i}^{\pm}\, =\, (E_{i}\,{\pm}\,p)/\sqrt{2}
  \label{kine-pipm},
  \end{equation}
  \begin{equation}
  k_{i}\, =\, x_{i}\,p_{i}+(0,0,\vec{k}_{i{\perp}})
  \label{kine-ki},
  \end{equation}
  \begin{equation}
 {\epsilon}_{1}^{\parallel}\, =\,
  \frac{p_{1}}{m_{1}}-\frac{m_{1}}{p_{1}{\cdot}n_{+}}n_{+}
  \label{kine13-longe},
  \end{equation}
  \begin{equation}
 {\epsilon}_{2}^{\parallel}\, =\,
  \frac{p_{2}}{m_{2}}-\frac{m_{2}}{p_{2}{\cdot}n_{-}}n_{-}
  \label{kine-longe},
  \end{equation}
  \begin{equation}
 {\epsilon}_{1,2}^{\perp}\, =\, (0,0,\vec{1})
  \label{kine-transe},
  \end{equation}
  \begin{equation}
  n_{+}=(1,0,0)
  \label{kine-nullp},
  \end{equation}
  \begin{equation}
  n_{-}=(0,1,0)
  \label{kine-nullm},
  \end{equation}
  \begin{equation}
  s\, =\, 2\,p_{2}{\cdot}p_{3}
  \label{kine-s},
  \end{equation}
  \begin{equation}
  t\, =\, 2\,p_{1}{\cdot}p_{2}\, =\ 2\,m_{1}\,E_{2}
  \label{kine-t},
  \end{equation}
  \begin{equation}
  u\, =\, 2\,p_{1}{\cdot}p_{3}\, =\ 2\,m_{1}\,E_{3}
  \label{kine-u},
  \end{equation}
  \begin{equation}
  p = \frac{\sqrt{ [m_{1}^{2}-(m_{2}+m_{3})^{2}]\,[m_{1}^{2}-(m_{2}-m_{3})^{2}] }}{2\,m_{1}}
  \label{kine-pcm},
  \end{equation}
  where $x_{i}$ and $\vec{k}_{i{\perp}}$ are the longitudinal momentum
  fraction and transverse momentum of valence quarks, respectively;
  ${\epsilon}_{i}^{\parallel}$ and ${\epsilon}_{i}^{\perp}$ are the
  longitudinal and transverse polarization vectors, respectively,
  and satisfy relations ${\epsilon}_{i}^{2}$ $=$ $-1$
  and ${\epsilon}_{i}{\cdot}p_{i}$ $=$ $0$;
  the subscript $i$ on variables $p_{i}$, $E_{i}$, $m_{i}$,
  ${\epsilon}_{i}$ corresponds to participating hadrons,
  namely, $i$ $=$ $1$ for ${\Upsilon}(nS)$ meson,
  $i$ $=$ $2$ for the recoiled $B_{c}^{\ast}$ meson,
  $i$ $=$ $3$ for the emitted pseudoscalar meson;
  $n_{+}$ and $n_{-}$ are positive and negative null vectors,
  respectively; $s$, $t$ and $u$ are the Lorentz-invariant
  variables; $p$ is the common momentum of final states.
  The notation of momentum is displayed in Fig. \ref{fig:fey}(a).

  \subsection{Wave functions}
  \label{sec0204}
  With the notation in \cite{jhep0605,jhep0703},
  wave functions are defined as
  \begin{equation}
 {\langle}0{\vert}b_{i}(z)\bar{b}_{j}(0){\vert}
 {\Upsilon}(p_{1},{\epsilon}_{1}^{{\parallel}}){\rangle}\,
 =\, \frac{f_{{\Upsilon}}}{4}{\int}dk_{1}\,e^{-ik_{1}{\cdot}z}
  \Big\{ \!\!\not{\epsilon}_{1}^{{\parallel}} \Big[
   m_{1}\,{\Phi}_{\Upsilon}^{v}(k_{1})
  -\!\!\not{p}_{1}\, {\Phi}_{\Upsilon}^{t}(k_{1})
  \Big] \Big\}_{ji}
  \label{wave-bb-long},
  \end{equation}
  \begin{equation}
 {\langle}0{\vert}b_{i}(z)\bar{b}_{j}(0){\vert}
 {\Upsilon}(p_{1},{\epsilon}_{1}^{{\perp}}){\rangle}\,
 =\, \frac{f_{{\Upsilon}}}{4}{\int}dk_{1}\,e^{-ik_{1}{\cdot}z}
  \Big\{ \!\!\not{\epsilon}_{1}^{{\perp}} \Big[
   m_{1}\,{\Phi}_{\Upsilon}^{V}(k_{1})
  -\!\!\not{p}_{1}\, {\Phi}_{\Upsilon}^{T}(k_{1})
  \Big] \Big\}_{ji}
  \label{wave-bb-perp},
  \end{equation}
  \begin{equation}
 {\langle}B_{c}^{\ast}(p_{2},{\epsilon}_{2}^{{\parallel}})
 {\vert}\bar{c}_{i}(z)b_{j}(0){\vert}0{\rangle}\ =\
  \frac{f_{B_{c}^{\ast}}}{4}{\int}_{0}^{1}dk_{3}\,e^{ik_{2}{\cdot}z}
  \Big\{ \!\not{\epsilon}_{2}^{{\parallel}} \Big[
   m_{2}\,{\Phi}_{B_{c}^{\ast}}^{v}(k_{2})
  +\!\not{p}_{2}\, {\Phi}_{B_{c}^{\ast}}^{t}(k_{2})
  \Big] \Big\}_{ji}
  \label{wave-bc-long},
  \end{equation}
  \begin{equation}
 {\langle}B_{c}^{{\ast}}(p_{2},{\epsilon}_{2}^{{\perp}})
 {\vert}\bar{c}_{i}(z)b_{j}(0){\vert}0{\rangle}\ =\
  \frac{f_{B_{c}^{\ast}}}{4}{\int}_{0}^{1}dk_{2}\,e^{ik_{2}{\cdot}z}
  \Big\{ \!\not{\epsilon}_{2}^{{\perp}} \Big[
   m_{2}\,{\Phi}_{B_{c}^{\ast}}^{V}(k_{2})
 +\!\not{p}_{2}\, {\Phi}_{B_{c}^{\ast}}^{T}(k_{2})
  \Big] \Big\}_{ji}
  \label{wave-bc-perp},
  \end{equation}
  \begin{eqnarray}
  & &
 {\langle}P(p_{3}){\vert}u_{i}(0)\bar{q}_{j}(z){\vert}0{\rangle}
  \nonumber \\ &=&
  \frac{i\,f_{P}}{4} {\int}dk_{3}\,e^{ik_{3}{\cdot}z}
  \Big\{ {\gamma}_{5}\Big[ \!\!\not{p}_{3}\,{\Phi}_{P}^{a}(k_{3})
  + {\mu}_{P}{\Phi}_{P}^{p}(k_{3})
  + {\mu}_{P}(\!\not{n}_{-}\!\!\not{n}_{+}\!-\!1)\,{\Phi}_{P}^{t}(k_{3})
  \Big] \Big\}_{ji}
  \label{wave-pim},
  \end{eqnarray}
  where $f_{{\Upsilon}}$, $f_{B_{c}^{\ast}}$, $f_{P}$ are
  decay constants of ${\Upsilon}(nS)$, $B_{c}^{\ast}$,
  $P$ mesons, respectively.

  Considering mass relations of $m_{{\Upsilon}(nS)}$ ${\simeq}$
  $2m_{b}$ and $m_{B_{c}^{\ast}}$ ${\simeq}$ $m_{b}$ $+$ $m_{c}$,
  it might assume that the motion of heavy valence quarks in
  ${\Upsilon}(nS)$ and $B_{c}^{\ast}$ mesons is nearly nonrelativistic.
  The wave functions of ${\Upsilon}(nS)$ and $B_{c}^{\ast}$
  mesons could be approximately described with nonrelativistic quantum
  chromodynamics (NRQCD) \cite{prd46,prd51,rmp77} and time-independent
  Schr\"{o}dinger equation.
  For an isotropic harmonic oscillator potential, the eigenfunctions of
  stationary state with quantum numbers $nL$ are written as \cite{plb751}
   \begin{equation}
  {\phi}_{1S}(\vec{k})\
  {\sim}\ e^{-\vec{k}^{2}/2{\beta}^{2}}
   \label{wave-p-1s},
   \end{equation}
   \begin{equation}
  {\phi}_{2S}(\vec{k})\ {\sim}\
   e^{-\vec{k}^{2}/2{\beta}^{2}}
   ( 2\vec{k}^{2}-3{\beta}^{2} )
   \label{wave-p-r2s},
   \end{equation}
   \begin{equation}
  {\phi}_{3S}(\vec{k})\ {\sim}\
   e^{-\vec{k}^{2}/2{\beta}^{2}}
   ( 4\vec{k}^{4}-20\vec{k}^{2}{\beta}^{2}+15{\beta}^{4} )
   \label{wave-p-r3s},
   \end{equation}
  where parameter ${\beta}$ determines the average
  transverse momentum, i.e.,
  ${\langle}nS{\vert}k^{2}_{\perp}{\vert}nS{\rangle}$
  ${\sim}$ ${\beta}^{2}$.
  Employing the substitution ansatz \cite{xiao},
   \begin{equation}
   \vec{k}^{2}\ {\to}\ \frac{1}{4} \sum\limits_{i}
   \frac{\vec{k}_{i\perp}^{2}+m_{q_{i}}^{2}}{x_{i}}
   \label{wave-kt},
   \end{equation}
  where $x_{i}$ and $m_{q_{i}}$ are the
  longitudinal momentum fraction and mass of valence quark,
  respectively,
  then integrating out $\vec{k}_{\perp}$ and combining with
  their asymptotic forms, the distribution
  amplitudes (DAs) for ${\Upsilon}(nS)$ and $B_{c}^{\ast}$
  mesons can be written as \cite{plb751},
   \begin{equation}
  {\phi}_{{\Upsilon}(1S)}^{v,T}(x) = A\, x\bar{x}\,
  {\exp}\Big\{ -\frac{m_{b}^{2}}{8\,{\beta}_{1}^{2}\,x\,\bar{x}} \Big\}
   \label{wave-1s-bblv},
   \end{equation}
   \begin{equation}
  {\phi}_{{\Upsilon}(1S)}^{t}(x) = B\, t^{2}\,
  {\exp}\Big\{ -\frac{m_{b}^{2}}{8\,{\beta}_{1}^{2}\,x\,\bar{x}} \Big\}
   \label{wave-1s-bblt},
   \end{equation}
   \begin{equation}
  {\phi}_{{\Upsilon}(1S)}^{V}(x) = C\, (1+t^{2})\,
  {\exp}\Big\{ -\frac{m_{b}^{2}}{8\,{\beta}_{1}^{2}\,x\,\bar{x}} \Big\}
   \label{wave-1s-bbtv},
   \end{equation}
   \begin{equation}
  {\phi}_{{\Upsilon}(2S)}^{v,t,V,T}(x) = D\,
  {\phi}_{{\Upsilon}(1S)}^{v,t,V,T}(x)\,
   \Big\{ 1+\frac{m_{b}^{2}}{2\,{\beta}_{1}^{2}\,x\,\bar{x}} \Big\}
   \label{wave-2s-bb},
   \end{equation}
   \begin{equation}
  {\phi}_{{\Upsilon}(3S)}^{v,t,V,T}(x) = E\,
  {\phi}_{{\Upsilon}(1S)}^{v,t,V,T}(x)\,
   \Big\{ \Big( 1-\frac{m_{b}^{2}}{2\,{\beta}_{1}^{2}\,x\,\bar{x}} \Big)^{2}
   +6 \Big\}
   \label{wave-3s-bb},
   \end{equation}
   \begin{equation}
  {\phi}_{B_{c}^{\ast}}^{v,T}(x) = F\, x\bar{x}\,
  {\exp}\Big\{ -\frac{\bar{x}\,m_{c}^{2}+x\,m_{b}^{2}}
                     {8\,{\beta}_{2}^{2}\,x\,\bar{x}} \Big\}
   \label{wave-bclv},
   \end{equation}
   \begin{equation}
  {\phi}_{B_{c}^{\ast}}^{t}(x) = G\, t^{2}\,
  {\exp}\Big\{ -\frac{\bar{x}\,m_{c}^{2}+x\,m_{b}^{2}}
                     {8\,{\beta}_{2}^{2}\,x\,\bar{x}} \Big\}
   \label{wave-bclt},
   \end{equation}
   \begin{equation}
  {\phi}_{B_{c}^{\ast}}^{V}(x) = H\, (1-t^{2})\,
  {\exp}\Big\{ -\frac{\bar{x}\,m_{c}^{2}+x\,m_{b}^{2}}
                     {8\,{\beta}_{2}^{2}\,x\,\bar{x}} \Big\}
   \label{wave-bctv},
   \end{equation}
   where $\bar{x}$ $=$ $1$ $-$ $x$; $t$ $=$ $x$ $-$ $\bar{x}$.
   According to NRQCD
   power counting rules \cite{prd46}, ${\beta}_{i}$ ${\simeq}$
   ${\xi}_{i}\,{\alpha}_{s}({\xi}_{i})$ with ${\xi}_{i}$ $=$
   $m_{i}/2$ and QCD coupling constant ${\alpha}_{s}$.
   The exponential function represents $k_{\perp}$ distribution.
   Parameters of $A$, $B$, $C$, $D$, $E$, $F$, $G$, $H$ are
   normalization coefficients satisfying with the conditions
   \begin{equation}
  {\int}_{0}^{1}dx\,{\phi}_{{\Upsilon}(nS)}^{i}(x) =
  {\int}_{0}^{1}dx\,{\phi}^{i}_{B_{c}^{\ast}}(x)=1
   \quad \text{for}\ \ i=v,t,V,T
   \label{normalization}.
   \end{equation}
  \begin{figure}[h]
  \includegraphics[width=0.95\textwidth,bb=75 630 530 720]{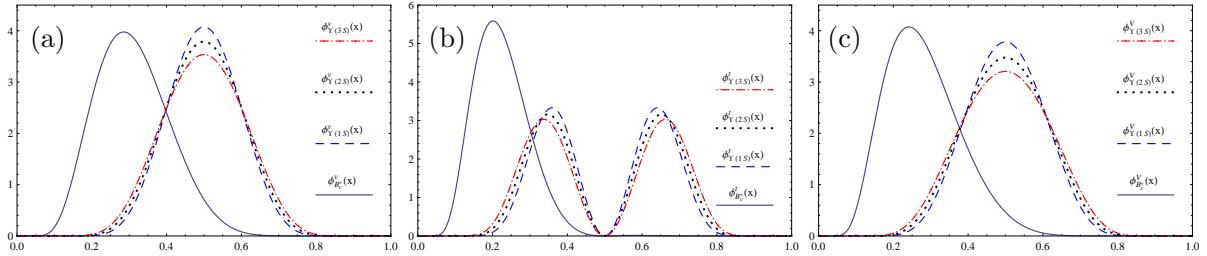}
  \caption{The normalized distribution amplitudes for
   ${\Upsilon}(nS)$ and $B_{c}^{\ast}$ mesons.}
  \label{fig:wave}
  \end{figure}

  The shape lines of normalized DAs for ${\Upsilon}(nS)$ and $B_{c}^{\ast}$
  mesons are showed in Fig. \ref{fig:wave}.
  It is clearly seen that
  (1) DAs for ${\Upsilon}(nS)$ and $B_{c}^{\ast}$ mesons fall quickly down
  to zero at endpoint $x$, $\bar{x}$ ${\to}$ $0$ due to suppression from
  exponential functions;
  (2) DAs for ${\Upsilon}(nS)$ meson are symmetric under the interchange
  of momentum fractions $x$ ${\leftrightarrow}$ $\bar{x}$, and DAs for
  $B_{c}^{\ast}$ meson are basically consistent with the feature that
  valence quarks share momentum fractions according to their masses.

  Our study shows that only the leading twist (twist-2) DAs of
  the emitted light pseudoscalar meson $P$ is involved in decay
  amplitudes (see Appendix \ref{blocks-01}).
  The twist-2 DAs has the expansion \cite{jhep0605}:
   \begin{equation}
  {\phi}_{P}^{a}(x)= 6\,x\,\bar{x}\, \sum\limits_{i=0}a_{i}\,C_{i}^{3/2}(t)
   \label{wave-pi-01},
   \end{equation}
  and are normalized as
   \begin{equation}
  {\int}_{0}^{1}{\phi}_{P}^{a}(x)\,dx= 1
   \label{wave-pi-02},
   \end{equation}
  where $C_{i}^{3/2}(t)$ are Gegenbauer polynomials,
   \begin{equation}
   C_{0}^{3/2}(t)=1, \quad
   C_{1}^{3/2}(t)=3\,t, \quad
   C_{2}^{3/2}(t)=\frac{3}{2}\,(5\,t^{2}-1), \quad
   \cdots
   \label{wave-p-c}
   \end{equation}
  and each term corresponds to a nonperturbative
  Gegenbauer moment $a_{i}$; note that $a_{0}$ $=$ $1$
  due to the normalization condition Eq.(\ref{wave-pi-02});
  the $G$-parity invariance of the pion DAs requires Gegenbauer
  moment $a_{i}$ $=$ $0$ for $i$ $=$ $1$, $3$, $5$ ${\cdots}$.

  \subsection{Decay amplitudes}
  \label{sec0205}
\begin{figure}[h]
  \includegraphics[width=0.95\textwidth,bb=75 620 530 730]{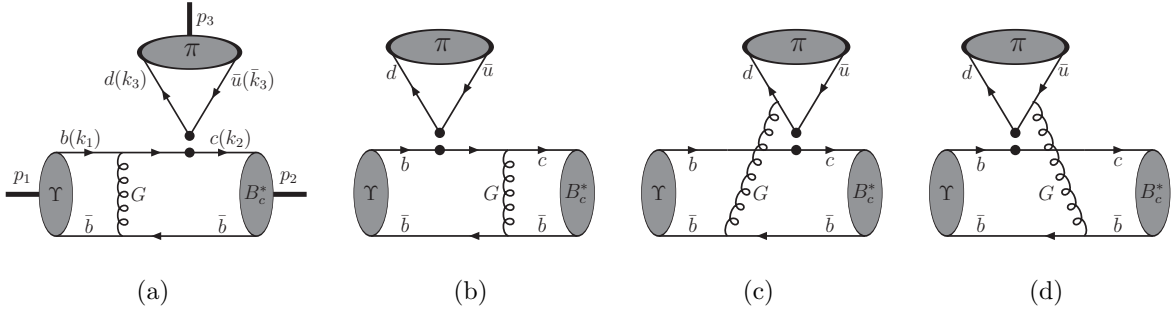}
  \caption{Feynman diagrams for ${\Upsilon}(nS)$ ${\to}$
  $B_{c}^{\ast}{\pi}$ decay with the pQCD approach, including
  factorizable emission diagrams (a,b) and nonfactorizable
  emission diagrams (c,d).}
  \label{fig:fey}
  \end{figure}

  The Feynman diagrams for ${\Upsilon}(nS)$ ${\to}$
  $B_{c}^{\ast}{\pi}$ weak decay are shown in Fig. \ref{fig:fey}.
  There are two types. One is factorizable emission topology
  where gluon attaches to quarks in the same meson, and the
  other is nonfactorizable emission topology where gluon connects
  to quarks between different mesons.

  With the pQCD master formula Eq.(\ref{hadronic}),
  the amplitude for ${\Upsilon}(nS)$ ${\to}$ $B_{c}^{\ast}P$
  decay can be expressed as \cite{prd66},
   \begin{equation}
  {\cal A}({\Upsilon}(nS){\to}B_{c}^{\ast}P)\ =\
  {\cal A}_{L}({\epsilon}_{1}^{{\parallel}},{\epsilon}_{2}^{{\parallel}})
 +{\cal A}_{N}({\epsilon}_{1}^{{\perp}},{\epsilon}_{2}^{{\perp}})
 +i\,{\cal A}_{T}\,{\varepsilon}_{{\mu}{\nu}{\alpha}{\beta}}\,
  {\epsilon}_{1}^{{\mu}}\,{\epsilon}_{2}^{{\nu}}\,
   p_{1}^{\alpha}\,p_{2}^{\beta}
   \label{eq:amp01},
   \end{equation}
  which is conventionally written as the helicity amplitudes \cite{prd66},
   \begin{equation}
  {\cal A}_{0}\ =\ -C_{\cal A}\, \sum\limits_{j}
  {\cal A}_{L}^{j}({\epsilon}_{1}^{{\parallel}},{\epsilon}_{2}^{{\parallel}})
   \label{eq:amp02},
   \end{equation}
   \begin{equation}
  {\cal A}_{\parallel}\ =\ \sqrt{2}\,C_{\cal A}\, \sum\limits_{j}
  {\cal A}_{N}^{j}({\epsilon}_{1}^{{\perp}},{\epsilon}_{2}^{{\perp}})
   \label{eq:amp03},
   \end{equation}
   \begin{equation}
  {\cal A}_{\perp}\ =\ \sqrt{2}\,C_{\cal A}\,m_{1}\,p\, \sum\limits_{j} {\cal A}_{T}^{j}
   \label{eq:amp04},
   \end{equation}
   \begin{equation}
  C_{\cal A}\ =\ i\, V_{cb} V_{uq}^{\ast}\,\frac{G_{F}}{\sqrt{2}}\,
  \frac{C_{F}}{N_{c}}\, {\pi}\, f_{{\Upsilon}}\,f_{B_{c}^{\ast}}\, f_{P}
   \label{eq:amp05},
   \end{equation}
  where $C_{F}$ $=$ $4/3$ and the color number $N_{c}$ $=$ $3$;
  the subscript $i$ on $A_{i}^{j}$ corresponds to three different
  helicity amplitudes, i.e., $i$ $=$ $L$, $N$, $T$;
  the superscript $j$ on $A_{i}^{j}$ denotes to indices of
  Fig. \ref{fig:fey}.
  The explicit expressions of building blocks ${\cal A}_{i}^{j}$
  are collected in Appendix \ref{blocks-01}.

  \section{Numerical results and discussion}
  \label{sec03}
  In the center-of-mass of ${\Upsilon}(nS)$ meson,
  branching ratio ${\cal B}r$ for ${\Upsilon}(nS)$ ${\to}$
  $B_{c}^{\ast}P$ decay are defined as
   \begin{equation}
  {\cal B}r\ =\ \frac{1}{12{\pi}}\,
   \frac{p}{m_{{\Upsilon}}^{2}{\Gamma}_{{\Upsilon}}}\, \Big\{
  {\vert}{\cal A}_{0}{\vert}^{2}+{\vert}{\cal A}_{\parallel}{\vert}^{2}
 +{\vert}{\cal A}_{\perp}{\vert}^{2} \Big\}
   \label{br}.
   \end{equation}

  The input parameters are listed in Table \ref{tab:bb} and \ref{tab:input}.
  If not specified explicitly, we will take their central
  values as the default inputs.
  Our numerical results are collected in Table. \ref{tab:case},
  where the first uncertainty comes from scale $(1{\pm}0.1)t_{i}$
  and the expression of $t_{i}$ is given in Eq.(\ref{tab}) and Eq.(\ref{tcd});
  the second uncertainty is from mass $m_{b}$ and $m_{c}$;
  the third uncertainty is from hadronic parameters including
  decay constants and Gegenbauer moments; the fourth uncertainty
  is from CKM parameters.
  The followings are some comments.

   \begin{table}[h]
   \caption{The numerical values of input parameters.}
   \label{tab:input}
   \begin{ruledtabular}
   \begin{tabular}{ll}
   \multicolumn{2}{c}{The Wolfenstein parameters} \\ \hline
     $A$ $=$ $0.814^{+0.023}_{-0.024}$ \cite{pdg},
   & ${\lambda}$ $=$ $0.22537{\pm}0.00061$ \cite{pdg}, \\ \hline
   \multicolumn{2}{c}{Mass, decay constant and Gegenbauer moments} \\ \hline
    $m_{b}$ $=$ $4.78{\pm}0.06$ GeV \cite{pdg},
  & $f_{\pi}$   $=$ $130.41{\pm}0.20$ MeV \cite{pdg}, \\
    $m_{c}$ $=$ $1.67{\pm}0.07$ GeV \cite{pdg},
  & $f_{K}$     $=$ $156.2{\pm}0.7$ MeV \cite{pdg}, \\
    $m_{B_{c}^{\ast}}$ $=$ $6332{\pm}9$ MeV \cite{mbc},
  & $f_{B_{c}^{\ast}}$ $=$ $422{\pm}13$ MeV \cite{fbc}\footnotemark[3], \\
    $a_{1}^{K}$ (1GeV) $=$ $-0.06{\pm}0.03$ \cite{jhep0605},
  & $f_{{\Upsilon}(1S)}$ $=$ $676.4{\pm}10.7$ MeV \cite{plb751}, \\
    $a_{2}^{K}$ (1GeV) $=$ $0.25{\pm}0.15$ \cite{jhep0605},
  & $f_{{\Upsilon}(2S)}$ $=$ $473.0{\pm}23.7$ MeV \cite{plb751}, \\
    $a_{2}^{\pi}$ (1GeV) $=$ $0.25{\pm}0.15$ \cite{jhep0605},
  & $f_{{\Upsilon}(3S)}$ $=$ $409.5{\pm}29.4$ MeV \cite{plb751}.
   \end{tabular}
   \end{ruledtabular}
   \end{table}
  \footnotetext[3]{The decay constant $f_{B_{c}^{\ast}}$ cannot be
  extracted from the experimental data because of no measurement
  on $B_{c}^{\ast}$ weak decay at the present time.
  Theoretically, the value of $f_{B_{c}^{\ast}}$ has been estimated,
  for example, in Ref. \cite{epja49} with the QCD sum rules.
  From Table. 3 of Ref. \cite{epja49}, one can see that the value
  of $f_{B_{c}^{\ast}}$ are model-dependent. In our calculation,
  we will take the latest value given by the lattice QCD approach
  \cite{fbc} just to offer an order of magnitude estimation on
  branching ratio for ${\Upsilon}(nS)$ ${\to}$ $B_{c}^{\ast}P$ decays.}

   \begin{table}[h]
   \caption{Branching ratio for ${\Upsilon}(nS)$ ${\to}$ $B_{c}^{\ast}P$ decays.}
   \label{tab:case}
   \begin{ruledtabular}
   \begin{tabular}{cccc}
   modes &
   ${\Upsilon}(1S)$ ${\to}$ $B_{c}^{\ast}{\pi}$ &
   ${\Upsilon}(2S)$ ${\to}$ $B_{c}^{\ast}{\pi}$ &
   ${\Upsilon}(3S)$ ${\to}$ $B_{c}^{\ast}{\pi}$  \\ \hline
   $10^{10}{\times}{\cal B}r$ &
   $4.35^{+0.29+0.19+0.44+0.17}_{-0.24-0.41-0.31-0.30}$ &
   $2.28^{+0.13+0.26+0.40+0.09}_{-0.03-0.35-0.16-0.15}$ &
   $2.14^{+0.12+0.09+0.48+0.07}_{-0.12-0.41-0.15-0.15}$ \\ \hline
   modes &
   ${\Upsilon}(1S)$ ${\to}$ $B_{c}^{\ast}K$ &
   ${\Upsilon}(2S)$ ${\to}$ $B_{c}^{\ast}K$ &
   ${\Upsilon}(3S)$ ${\to}$ $B_{c}^{\ast}K$ \\ \hline
   $10^{11}{\times}{\cal B}r$ &
   $3.45^{+0.23+0.13+0.38+0.13}_{-0.21-0.35-0.27-0.25}$ &
   $1.91^{+0.11+0.07+0.36+0.07}_{-0.09-0.31-0.15-0.14}$ &
   $1.65^{+0.09+0.08+0.40+0.05}_{-0.21-0.33-0.13-0.12}$
   \end{tabular}
   \end{ruledtabular}
   \end{table}

  (1)
  Branching ratio for ${\Upsilon}(nS)$ ${\to}$ $B_{c}^{\ast}{\pi}$
  decay is about ${\cal O}(10^{-10})$ with pQCD approach,
  which is well within the measurement potential of LHC and SuperKEKB.
  For example, experimental studies have showed that production cross
  sections for ${\Upsilon}(nS)$ meson in p-p and p-Pb collisions
  are a few ${\mu}b$ at the LHCb \cite{epjc74lhcb,jhep1407} and
  ALICE \cite{prd87,plb740} detectors.
  Consequently, there will be more than $10^{12}$ ${\Upsilon}(nS)$
  data samples per $ab^{-1}$ data collected by the LHCb and ALICE,
  corresponding to a few hundreds of ${\Upsilon}(nS)$ ${\to}$
  $B_{c}^{\ast}{\pi}$ events.
  Branching ratio for ${\Upsilon}(nS)$ ${\to}$ $B_{c}^{\ast}K$ decay,
  ${\cal O}(10^{-11})$, is generally less than that for ${\Upsilon}(nS)$
  ${\to}$ $B_{c}^{\ast}{\pi}$ decay by one order of magnitude due to
  the CKM suppression, ${\vert}V_{us}^{\ast}/V_{ud}^{\ast}{\vert}^{2}$
  ${\sim}$ ${\lambda}^{2}$.

  \begin{figure}[h]
  \includegraphics[width=0.99\textwidth,bb=85 625 540 720]{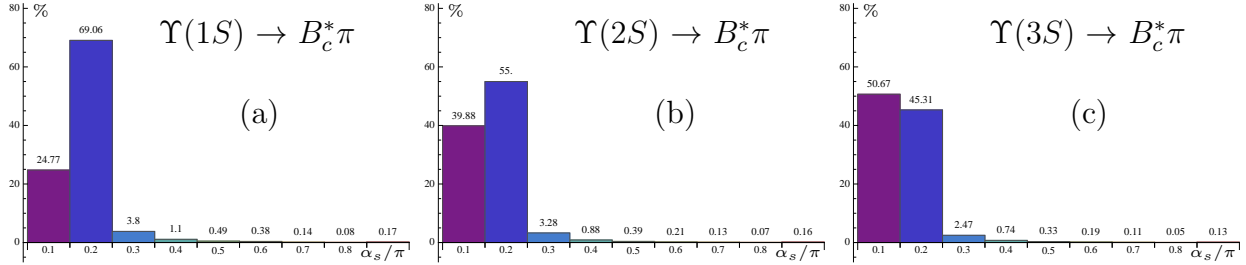}
  \caption{The contributions to branching ratios for
  ${\Upsilon}(1S)$ ${\to}$ $B_{c}^{\ast}{\pi}$ decay (a),
  ${\Upsilon}(2S)$ ${\to}$ $B_{c}^{\ast}{\pi}$ decay (b) and
  ${\Upsilon}(3S)$ ${\to}$ $B_{c}^{\ast}{\pi}$ decay (c)
  from different
  region of ${\alpha}_{s}/{\pi}$ (horizontal axises), where the
  numbers over histogram denote the percentage of the corresponding
  contributions.}
  \label{fig:br-as}
  \end{figure}

  (2)
  As it is well known, due to the large mass of $B_{c}^{\ast}$,
  the momentum transition in the ${\Upsilon}(nS)$ ${\to}$
  $B_{c}^{\ast}P$ decay may be not large enough. One might
  naturally wonder whether the pQCD approach is applicable
  and whether the perturbative calculation is reliable.
  Therefore, it is necessary to check what percentage
  of the contributions comes from the perturbative region.
  The contributions to branching ratio for ${\Upsilon}(nS)$
  ${\to}$ $B_{c}^{\ast}{\pi}$ decay from different
  ${\alpha}_{s}/{\pi}$ region are showed in Fig. \ref{fig:br-as}.
  It can be clearly seen that more than 93\% (97\%) contributions
  come from the ${\alpha}_{s}/{\pi}$ ${\le}$ $0.2$ (0.3) region,
  implying that the ${\Upsilon}(nS)$ ${\to}$ $B_{c}^{\ast}{\pi}$
  decay is computable with the pQCD approach.
  As the discussion in \cite{pqcd1,pqcd2,pqcd3},
  there are many factors for this, for example,
  the choice of the typical scale, retaining the quark transverse
  moment and introducing the Sudakov factor to suppress the
  nonperturbative contributions, which deserve much attention
  and further investigation.

  (3)
  Because of the relations among masses $m_{{\Upsilon}(3S)}$
  $>$ $m_{{\Upsilon}(2S)}$ $>$ $m_{{\Upsilon}(1S)}$ resulting
  in the fact that phase space increases with the radial
  quantum number $n$, in addition, the relations
  among decay widths ${\Gamma}_{{\Upsilon}(3S)}$ $<$
  ${\Gamma}_{{\Upsilon}(2S)}$ $<$ ${\Gamma}_{{\Upsilon}(1S)}$,
  in principle, there should be relations among branching ratios
  ${\cal B}r({\Upsilon}(3S){\to}B_{c}^{\ast}P)$
  $>$ ${\cal B}r({\Upsilon}(2S){\to}B_{c}^{\ast}P)$ $>$
  ${\cal B}r({\Upsilon}(1S){\to}B_{c}^{\ast}P)$ for the same
  pseudoscalar meson $P$.
  But the numerical results in Table. \ref{tab:case} are
  beyond such expectation. Why?
  The reason is that the factor of $p/m_{{\Upsilon}(nS)}^{2}$
  in Eq.(\ref{br}) has almost the same value for $n$ ${\le}$ $3$,
  so branching ratio is proportional to factor
  $f_{{\Upsilon}(nS)}^{2}/{\Gamma}_{{\Upsilon}(nS)}$ with
  the maximal value $f_{{\Upsilon}(1S)}^{2}/{\Gamma}_{{\Upsilon}(1S)}$
  for $n$ ${\le}$ $3$.
  Besides, contributions from ${\alpha}_{s}/{\pi}$ ${\in}$ $[0.2,0.3]$
  regions decrease with $n$ (see Fig. \ref{fig:br-as}), which
  enhance the decay amplitudes.

  (4)
  Besides the uncertainties listed in Table \ref{tab:case},
  other factors, such as the models of wave functions,
  contributions of higher order corrections to HME,
  relativistic effects, and so on, deserve
  the dedicated study. Our results just provide an order
  of magnitude estimation.

  \section{Summary}
  \label{sec04}
  The ${\Upsilon}(nS)$ decay via the weak interaction, as a
  complementary to strong and electromagnetic decay mechanism,
  is allowable within the standard model.
  Based on the potential prospects of ${\Upsilon}(nS)$ physics
  at high-luminosity collider experiment,
  ${\Upsilon}(nS)$ decay into $B_{c}^{\ast}{\pi}$ and
  $B_{c}^{\ast}K$ final states is investigated with the pQCD
  approach firstly.
  It is found that
  (1) the dominant contributions come from perturbative regions
  ${\alpha}_{s}/{\pi}$ ${\le}$ $0.3$, which might imply that
  the pQCD calculation is practicable and workable;
  (2) there is a promiseful possibility of searching for
  ${\Upsilon}(nS)$ ${\to}$ $B_{c}^{\ast}{\pi}$ ($B_{c}^{\ast}K$)
  decay with branching ratio about $10^{-10}$ ($10^{-11}$)
  at the future experiments.

  \section*{Acknowledgments}
  The work is supported by the National Natural Science Foundation
  of China (Grant Nos. 11547014, 11475055, U1332103 and 11275057).

  \begin{appendix}
  \section{Building blocks for ${\Upsilon}$ ${\to}$ $B_{c}^{\ast}P$ decays}
  \label{blocks-01}
  The building blocks ${\cal A}_{i}^{j}$, where
  the superscript $j$ corresponds to indices of Fig. \ref{fig:fey} and
  the subscript $i$ relates with different helicity amplitudes,
  are expressed as follows.
   \begin{eqnarray}
  {\cal A}_{L}^{a} &=&
  {\int}_{0}^{1}dx_{1} {\int}_{0}^{1}dx_{2}
  {\int}_{0}^{\infty}b_{1} db_{1}
  {\int}_{0}^{\infty}b_{2} db_{2}\,
  H_{ab}({\alpha}_{e},{\beta}_{a},b_{1},b_{2})\,
  E_{ab}(t_{a})\,{\phi}_{\Upsilon}^{v}(x_{1})\,
  {\alpha}_{s}(t_{a})
   \nonumber \\ & &
   a_{1}(t_{a})\, \Big\{
  {\phi}_{B_{c}^{\ast}}^{v}(x_{2})\, \Big[ m_{1}^{2}\,s
  -(4\,m_{1}^{2}\,p^{2}+m_{2}^{2}\,u)\,\bar{x}_{2} \Big]
 +{\phi}_{B_{c}^{\ast}}^{t}(x_{2})\,m_{2}\,m_{b}\,u \Big\}
   \label{figaL-LL},
   \end{eqnarray}
   \begin{eqnarray}
  {\cal A}_{N}^{a} &=&
  {\int}_{0}^{1}dx_{1} {\int}_{0}^{1}dx_{2}
  {\int}_{0}^{\infty}b_{1} db_{1}
  {\int}_{0}^{\infty}b_{2} db_{2}\,
  H_{ab}({\alpha}_{e},{\beta}_{a},b_{1},b_{2})\, E_{ab}(t_{a})\,
  {\phi}_{\Upsilon}^{V}(x_{1})
   \nonumber \\ & &
  {\alpha}_{s}(t_{a})\,  a_{1}(t_{a})\, m_{1}\, \Big\{
  {\phi}_{B_{c}^{\ast}}^{V}(x_{2})\,m_{2}\,(u-s\,\bar{x}_{2})
 +{\phi}_{B_{c}^{\ast}}^{T}(x_{2})\,m_{b}\,s \Big\}
   \label{figaN-LL},
   \end{eqnarray}
  \begin{eqnarray}
  {\cal A}_{T}^{a} &=& -2\,m_{1}\,
  {\int}_{0}^{1}dx_{1} {\int}_{0}^{1}dx_{2}
  {\int}_{0}^{\infty}b_{1} db_{1}
  {\int}_{0}^{\infty}b_{2} db_{2}\,
  H_{ab}({\alpha}_{e},{\beta}_{a},b_{1},b_{2})\,
  E_{ab}(t_{a})
   \nonumber \\ & &
  {\phi}_{\Upsilon}^{V}(x_{1})\,
  {\alpha}_{s}(t_{a})\,  a_{1}(t_{a})\, \Big\{
  {\phi}_{B_{c}^{\ast}}^{T}(x_{2})\,m_{b}
 +{\phi}_{B_{c}^{\ast}}^{V}(x_{2})\,m_{2}\,x_{2}
   \Big\}
   \label{figaT-LL},
   \end{eqnarray}
   \begin{eqnarray}
  {\cal A}_{L}^{b} &=&
  {\int}_{0}^{1}dx_{1} {\int}_{0}^{1}dx_{2}
  {\int}_{0}^{\infty}b_{1} db_{1}
  {\int}_{0}^{\infty}b_{2} db_{2}\,
  H_{ab}({\alpha}_{e},{\beta}_{b},b_{2},b_{1})\,
  E_{ab}(t_{b})\, {\phi}_{B_{c}^{\ast}}^{v}(x_{2})
   \nonumber \\ & &
  {\alpha}_{s}(t_{b})\,  a_{1}(t_{b})\,
   \Big\{ {\phi}_{\Upsilon}^{v}(x_{1})\, \Big[
  m_{2}^{2}\,u-m_{1}^{2}\,(s-4\,p^{2})\,\bar{x}_{1} \Big]
 +{\phi}_{\Upsilon}^{t}(x_{1})\, m_{1}\,m_{c}\,s \Big\}
   \label{figbL-LL},
   \end{eqnarray}
   \begin{eqnarray}
  {\cal A}_{N}^{b} &=&
  {\int}_{0}^{1}dx_{1} {\int}_{0}^{1}dx_{2}
  {\int}_{0}^{\infty}b_{1} db_{1}
  {\int}_{0}^{\infty}b_{2} db_{2}\,
  H_{ab}({\alpha}_{e},{\beta}_{b},b_{2},b_{1})\,
  E_{ab}(t_{b})\, {\phi}_{B_{c}^{\ast}}^{V}(x_{2})
   \nonumber \\ & &
 {\alpha}_{s}(t_{b})\,  a_{1}(t_{b})\,
  m_{2}\, \Big\{
  {\phi}_{\Upsilon}^{V}(x_{1})\,m_{1}\, (s-u\,\bar{x}_{1})
 +{\phi}_{\Upsilon}^{T}(x_{1})\,m_{c}\,u \Big\}
   \label{figbN-LL},
   \end{eqnarray}
   \begin{eqnarray}
  {\cal A}_{T}^{b} &=& - 2\,m_{2}
  {\int}_{0}^{1}dx_{1} {\int}_{0}^{1}dx_{2}
  {\int}_{0}^{\infty}b_{1} db_{1}
  {\int}_{0}^{\infty}b_{2} db_{2}\,
  H_{ab}({\alpha}_{e},{\beta}_{b},b_{2},b_{1})\,
  E_{ab}(t_{b})
   \nonumber \\ & &
  {\phi}_{B_{c}^{\ast}}^{V}(x_{2})\, {\alpha}_{s}(t_{b})\,
  a_{1}(t_{b})\, \Big\{
  {\phi}_{\Upsilon}^{V}(x_{1})\,m_{1}\,x_{1}
 +{\phi}_{\Upsilon}^{T}(x_{1})\,m_{c} \Big\}
   \label{figbT-LL-d},
   \end{eqnarray}
   \begin{eqnarray}
  {\cal A}_{L}^{c} &=& \frac{1}{N_{c}}
  {\int}_{0}^{1}dx_{1} {\int}_{0}^{1}dx_{2} {\int}_{0}^{1}dx_{3}
  {\int}_{0}^{\infty}db_{1}  {\int}_{0}^{\infty}b_{2}db_{2}
  {\int}_{0}^{\infty}b_{3}db_{3}\,
  H_{cd}({\alpha}_{e},{\beta}_{c},b_{2},b_{3})
   \nonumber \\ & &
  E_{cd}(t_{c})\, {\phi}_{P}^{a}(x_{3})\,
  {\alpha}_{s}(t_{c})\, C_{2}(t_{c})\, \Big\{
  {\phi}_{\Upsilon}^{v}(x_{1})\,
  {\phi}_{B_{c}^{\ast}}^{v}(x_{2})\,
  4\,m_{1}^{2}\,p^{2}\,(x_{1}-\bar{x}_{3})
   \nonumber \\ & &
   + {\phi}_{\Upsilon}^{t}(x_{1})\,
  {\phi}_{B_{c}^{\ast}}^{t}(x_{2})\,m_{1}\,m_{2}\,
  (u\,x_{1}-s\,x_{2}-2\,m_{3}^{2}\,\bar{x}_{3})
   \Big\} {\delta}(b_{1}-b_{2})
   \label{figcL-LL},
   \end{eqnarray}
   \begin{eqnarray}
  {\cal A}_{N}^{c} &=& \frac{1}{N_{c}}
  {\int}_{0}^{1}dx_{1} {\int}_{0}^{1}dx_{2} {\int}_{0}^{1}dx_{3}
  {\int}_{0}^{\infty}db_{1}  {\int}_{0}^{\infty}b_{2}db_{2}
  {\int}_{0}^{\infty}b_{3}db_{3}\,
  H_{cd}({\alpha}_{e},{\beta}_{c},b_{2},b_{3})\,
  E_{cd}(t_{c})\, C_{2}(t_{c})
   \nonumber \\ & &
  {\alpha}_{s}(t_{c})\,{\delta}(b_{1}-b_{2})\,
  {\phi}_{\Upsilon}^{T}(x_{1})\,
  {\phi}_{B_{c}^{\ast}}^{T}(x_{2})\,
  {\phi}_{P}^{a}(x_{3})\,
   \Big\{ m_{1}^{2}\,s\,(x_{1}-\bar{x}_{3})
   +m_{2}^{2}\,u\,(\bar{x}_{3}-x_{2}) \Big\}
   \label{figcN-LL},
   \end{eqnarray}
   \begin{eqnarray}
  {\cal A}_{T}^{c} &=& \frac{2}{N_{c}}
  {\int}_{0}^{1}dx_{1} {\int}_{0}^{1}dx_{2} {\int}_{0}^{1}dx_{3}
  {\int}_{0}^{\infty}db_{1}  {\int}_{0}^{\infty}b_{2}db_{2}
  {\int}_{0}^{\infty}b_{3}db_{3}\,
  H_{cd}({\alpha}_{e},{\beta}_{c},b_{2},b_{3})\,
  E_{cd}(t_{c})\, C_{2}(t_{c})
   \nonumber \\ & &
  {\alpha}_{s}(t_{c})\, {\delta}(b_{1}-b_{2})\,
  {\phi}_{\Upsilon}^{T}(x_{1})\,
  {\phi}_{B_{c}^{\ast}}^{T}(x_{2})\,
  {\phi}_{P}^{a}(x_{3})\,
   \Big\{ m_{1}^{2}\,(\bar{x}_{3}-x_{1})
  +m_{2}^{2}\,(x_{2}-\bar{x}_{3}) \Big\}
   \label{figcT-LL},
   \end{eqnarray}
   \begin{eqnarray}
  {\cal A}_{L}^{d} &=& \frac{1}{N_{c}}
  {\int}_{0}^{1}dx_{1} {\int}_{0}^{1}dx_{2} {\int}_{0}^{1}dx_{3}
  {\int}_{0}^{\infty}db_{1}  {\int}_{0}^{\infty}b_{2}db_{2}
  {\int}_{0}^{\infty}b_{3}db_{3}\,
  H_{cd}({\alpha}_{e},{\beta}_{d},b_{2},b_{3})
   \nonumber \\ & &
   E_{cd}(t_{d})\, {\phi}_{P}^{a}(x_{3})\,
  {\alpha}_{s}(t_{d})\, C_{2}(t_{d})\,
   \Big\{ {\phi}_{\Upsilon}^{v}(x_{1})\,
  {\phi}_{B_{c}^{\ast}}^{v}(x_{2})\,
  4\,m_{1}^{2}\,p^{2}\,(x_{3}-x_{2})
   \nonumber \\ & &
 +{\phi}_{\Upsilon}^{t}(x_{1})\,
  {\phi}_{B_{c}^{\ast}}^{t}(x_{2})\,
   m_{1}\, m_{2}\, (s\,x_{2} +2\,m_{3}^{2}\,x_{3}-u\,x_{1})
   \Big\}\, {\delta}(b_{1}-b_{2})
   \label{figdL-LL-p},
   \end{eqnarray}
   \begin{eqnarray}
  {\cal A}_{N}^{d} &=& \frac{1}{N_{c}}
  {\int}_{0}^{1}dx_{1} {\int}_{0}^{1}dx_{2} {\int}_{0}^{1}dx_{3}
  {\int}_{0}^{\infty}db_{1}  {\int}_{0}^{\infty}b_{2}db_{2}
  {\int}_{0}^{\infty}b_{3}db_{3}\,
  H_{cd}({\alpha}_{e},{\beta}_{d},b_{2},b_{3})\,
  E_{cd}(t_{d})\, C_{2}(t_{d})
   \nonumber \\ & &
  {\alpha}_{s}(t_{d})\, {\delta}(b_{1}-b_{2})\,
  {\phi}_{\Upsilon}^{T}(x_{1})\,
  {\phi}_{B_{c}^{\ast}}^{T}(x_{2})\,
  {\phi}_{P}^{a}(x_{3})\, \Big\{
   m_{1}^{2}\,s\,(x_{3}-x_{1})
  +m_{2}^{2}\,u\,(x_{2}-x_{3}) \Big\}
   \label{figdN-LL-p},
   \end{eqnarray}
   \begin{eqnarray}
  {\cal A}_{T}^{d} &=& \frac{2}{N_{c}}
  {\int}_{0}^{1}dx_{1} {\int}_{0}^{1}dx_{2} {\int}_{0}^{1}dx_{3}
  {\int}_{0}^{\infty}db_{1}  {\int}_{0}^{\infty}b_{2}db_{2}
  {\int}_{0}^{\infty}b_{3}db_{3}\,
  H_{cd}({\alpha}_{e},{\beta}_{d},b_{2},b_{3})\,
  E_{cd}(t_{d})\, C_{2}(t_{d})
   \nonumber \\ & &
  {\alpha}_{s}(t_{d})\, {\delta}(b_{1}-b_{2})\,
  {\phi}_{\Upsilon}^{T}(x_{1})\,
  {\phi}_{B_{c}^{\ast}}^{T}(x_{2})\,
  {\phi}_{P}^{a}(x_{3})\,
   \Big\{ m_{1}^{2}\,(x_{1}-x_{3})
         -m_{2}^{2}\,(x_{2}-x_{3}) \Big\}
   \label{figdT-LL-p},
   \end{eqnarray}
  where $\bar{x}_{i}$ $=$ $1$ $-$ $x_{i}$;
  variable $x_{i}$ is the longitudinal momentum fraction
  of the valence quark;
  $b_{i}$ is the conjugate variable of the
  transverse momentum $k_{i{\perp}}$;
  and ${\alpha}_{s}(t)$ is the QCD coupling at the
  scale of $t$; $a_{1}$ $=$ $C_{1}$ $+$ $C_{2}/N_{c}$.

  The function $H_{i}$ are defined as follows \cite{plb751}.
   \begin{eqnarray}
   H_{ab}({\alpha}_{e},{\beta},b_{i},b_{j})
   &=& K_{0}(\sqrt{-{\alpha}}b_{i})
   \Big\{ {\theta}(b_{i}-b_{j})
   K_{0}(\sqrt{-{\beta}}b_{i})
   I_{0}(\sqrt{-{\beta}}b_{j})
   + (b_{i}{\leftrightarrow}b_{j}) \Big\}
   \label{hab}, \\
   H_{cd}({\alpha}_{e},{\beta},b_{2},b_{3}) &=&
   \Big\{ {\theta}(-{\beta}) K_{0}(\sqrt{-{\beta}}b_{3})
  +\frac{{\pi}}{2} {\theta}({\beta}) \Big[
   iJ_{0}(\sqrt{{\beta}}b_{3})
   -Y_{0}(\sqrt{{\beta}}b_{3}) \Big] \Big\}
   \nonumber \\ &{\times}&
   \Big\{ {\theta}(b_{2}-b_{3})
   K_{0}(\sqrt{-{\alpha}}b_{2})
   I_{0}(\sqrt{-{\alpha}}b_{3})
   + (b_{2}{\leftrightarrow}b_{3}) \Big\}
   \label{hcd}
   \end{eqnarray}
  where $J_{0}$ and $Y_{0}$ ($I_{0}$ and $K_{0}$) are the
  (modified) Bessel function of the first and second kind,
  respectively;
  ${\alpha}_{e}$ (${\alpha}_{a}$) is the
  gluon virtuality of the emission (annihilation)
  topological diagrams;
  the subscript of the quark virtuality ${\beta}_{i}$
  corresponds to the indices of Fig. \ref{fig:fey}.
  The definition of the particle virtuality is
  listed as follows \cite{plb751}.
   \begin{eqnarray}
  {\alpha} &=& \bar{x}_{1}^{2}\,m_{1}^{2}
                +  \bar{x}_{2}^{2}\,m_{2}^{2}
                -  \bar{x}_{1}\,\bar{x}_{2}\,t
   \label{gluon-q2-e}, \\
  {\beta}_{a} &=& m_{1}^{2} - m_{b}^{2}
               +  \bar{x}_{2}^{2}\,m_{2}^{2}
               -  \bar{x}_{2}\,t
   \label{beta-fa}, \\
  {\beta}_{b} &=& m_{2}^{2} - m_{c}^{2}
               +  \bar{x}_{1}^{2}\,m_{1}^{2}
               -  \bar{x}_{1}\,t
   \label{beta-fb}, \\
  {\beta}_{c} &=& x_{1}^{2}\,m_{1}^{2}
               +  x_{2}^{2}\,m_{2}^{2}
               +  \bar{x}_{3}^{2}\,m_{3}^{2}
   \nonumber \\ &-&
                  x_{1}\,x_{2}\,t
               -  x_{1}\,\bar{x}_{3}\,u
               +  x_{2}\,\bar{x}_{3}\,s
   \label{beta-fc}, \\
  {\beta}_{d} &=& x_{1}^{2}\,m_{1}^{2}
               +  x_{2}^{2}\,m_{2}^{2}
               +  x_{3}^{2}\,m_{3}^{2}
    \nonumber \\ &-&
                  x_{1}\,x_{2}\,t
               -  x_{1}\,x_{3}\,u
               +  x_{2}\,x_{3}\,s
   \label{beta-fd}.
   \end{eqnarray}

  The typical scale $t_{i}$ and the Sudakov factor $E_{i}$
  are defined as follows, where the subscript $i$ corresponds
  to the indices of Fig. \ref{fig:fey}.
   \begin{eqnarray}
   t_{a(b)} &=& {\max}(\sqrt{-{\alpha}},\sqrt{-{\beta}_{a(b)}},1/b_{1},1/b_{2})
   \label{tab}, \\
   t_{c(d)} &=& {\max}(\sqrt{-{\alpha}},\sqrt{{\vert}{\beta}_{c(d)}{\vert}},1/b_{2},1/b_{3})
   \label{tcd},
   \end{eqnarray}
   \begin{eqnarray}
   E_{ab}(t) &=& {\exp}\{ -S_{{\Upsilon}}(t)-S_{B_{c}^{\ast}}(t) \}
   \label{sudakov-ab}, \\
   E_{cd}(t) &=& {\exp}\{ -S_{{\Upsilon}}(t)-S_{B_{c}^{\ast}}(t)-S_{P}(t) \}
   \label{sudakov-cd},
   \end{eqnarray}
   \begin{equation}
   S_{{\Upsilon}}(t)\ =\
   s(x_{1},p_{1}^{+},1/b_{1})
  +2{\int}_{1/b_{1}}^{t}\frac{d{\mu}}{\mu}{\gamma}_{q}
   \label{sudakov-bb},
   \end{equation}
   \begin{equation}
   S_{B_{c}^{\ast}}(t)\ =\
   s(x_{2},p_{2}^{+},1/b_{2})
  +2{\int}_{1/b_{2}}^{t}\frac{d{\mu}}{\mu}{\gamma}_{q}
   \label{sudakov-bc},
   \end{equation}
   \begin{equation}
   S_{{\pi},K}(t)\ =\
   s(x_{3},p_{3}^{+},1/b_{3})+s(\bar{x}_{3},p_{3}^{+},1/b_{3})
  +2{\int}_{1/b_{3}}^{t}\frac{d{\mu}}{\mu}{\gamma}_{q}
   \label{sudakov-pi-k},
   \end{equation}
  where ${\gamma}_{q}$ $=$ $-{\alpha}_{s}/{\pi}$ is the
  quark anomalous dimension;
  the explicit expression of $s(x,Q,1/b)$ can be found in
  the appendix of Ref. \cite{pqcd1}.
  \end{appendix}

  
  \end{document}